\documentclass[12pt,hyper]{JHEP3}
\usepackage{graphics,psfrag}
\usepackage{amsmath,amsthm,amssymb,epsfig,euscript,array,cite,cancel}
\usepackage{cases,empheq}
\theoremstyle{definition}

\theoremstyle{remark}

\newcounter{multieqs}

\newcommand{\be}{\begin{equation}}
\newcommand{\ee}{\end{equation}}
\newcommand{\eq}[1]{(\ref{#1})}
\newcommand{\bit}{\begin{itemize}}  \newcommand{\eit}{\end{itemize}}

\newcommand{\bm}[1]{\mbox{\boldmath $#1$}}
\newcommand{\rf}[1]{(\ref{#1})}

\def\bd{\begin{document}}
\def\ed{\end{document}}
\def\nn{\nonumber}
\def\bea{\begin{eqnarray}}
\def\eea{\end{eqnarray}}
\let\bm=\bibitem

\def\la{\langle}
\def\ra{\rangle}

\def\npb#1#2#3{Nucl. Phys. {\bf{B#1}} #3 (#2)}
\def\plb#1#2#3{Phys. Lett. {\bf{#1B}} #3 (#2)}
\def\prl#1#2#3{Phys. Rev. Lett. {\bf{#1}} #3 (#2)}
\def\prd#1#2#3{Phys. Rev. {D \bf{#1}} #3 (#2)}
\def\cmp#1#2#3{Comm. Math. Phys. {\bf{#1}} #3 (#2)}
\def\cqg#1#2#3{Class. Quantum Grav. {\bf{#1}} #3 (#2)}
\def\nppsa#1#2#3{Nucl. Phys. B (Proc. Suppl.) {\bf{#1A}}#3 (#2)}
\def\ap#1#2#3{Ann. of Phys. {\bf{#1}} #3 (#2)}
\def\ijmp#1#2#3{Int. J. Mod. Phys. {\bf{A#1}} #3 (#2)}
\def\rmp#1#2#3{Rev. Mod. Phys. {\bf{#1}} #3 (#2)}
\def\mpla#1#2#3{Mod. Phys. Lett. {\bf A#1} #3 (#2)}
\def\jhep#1#2#3{J. High Energy Phys. {\bf #1} #3 (#2)}
\def\atmp#1#2#3{Adv. Theor. Math. Phys. {\bf #1} #3 (#2)}

\def\N{{\cal N}}
\def\sst{\scriptscriptstyle}
\def\thetabar{\bar\theta}
\def\Tr{{\rm Tr}}
\def\one{\mbox{1 \kern-.59em {\rm l}}}

\def\a{\alpha}      \def\da{{\dot\alpha}}  \def\dA{{\dot A}}
\def\b{\beta}       \def\db{{\dot\beta}}  
\def\g{\gamma}  \def\G{\Gamma}  \def\dc{{\dot\gamma}}  
\def\d{\delta}  \def\D{\Delta}  \def\ddt{\dot\delta}  
\def\e{\epsilon}        \def\ve{\varepsilon}  
\def\f{\phi}    \def\F{\Phi}    \def\vvf{\f}  
\def\h{\eta}  
\def\k{\kappa}  
\def\l{\lambda} \def\L{\Lambda}  
\def\m{\mu} \def\n{\nu}  
\def\o{\omega}  
\def\p{\pi} \def\P{\Pi}  
\def\r{\rho}  
\def\s{\sigma}  \def\S{\Sigma}  
\def\t{\tau}  
\def\th{\theta} \def\Th{\Theta} \def\vth{\vartheta}  
\def\X{\Xeta}  
\def\z{\zeta}  

\def\na{\nabla}  
\def\cA{{\cal A}} \def\cB{{\cal B}} \def\cC{{\cal C}}  
\def\cD{{\cal D}} \def\cE{{\cal E}} \def\cF{{\cal F}}  
\def\cG{{\cal G}} \def\cH{{\cal H}} \def\cI{{\cal I}}  
\def\cJ{{\cal J}} \def\cK{{\cal K}} \def\cL{{\cal L}}  
\def\cM{{\cal M}} \def\cN{{\cal N}} \def\cO{{\cal O}}  
\def\cP{{\cal P}} \def\cQ{{\cal Q}} \def\cR{{\cal R}}  
\def\cS{{\cal S}} \def\cT{{\cal T}} \def\cU{{\cal U}}  
\def\cV{{\cal V}} \def\cW{{\cal W}} \def\cX{{\cal X}}  
\def\cY{{\cal Y}} \def\cZ{{\cal Z}}

\def\ua{\underline{\alpha}}  
\def\uc{\underline{\phantom{\alpha}}\!\!\!\gamma}  
\def\um{\underline{\mu}}  
\def\ud{\underline\delta}  
\def\ue{\underline\epsilon}  
\def\una{\underline a}\def\unA{\underline A}  
\def\unb{\underline b}\def\unB{\underline B}  
\def\unc{\underline c}\def\unC{\underline C}  
\def\und{\underline d}\def\unD{\underline D}  
\def\une{\underline e}\def\unE{\underline E}  
\def\unf{\underline{\phantom{e}}\!\!\!\! f}\def\unF{\underline F}  
\def\unm{\underline m}\def\unM{\underline M}  
\def\unn{\underline n}\def\unN{\underline N}  
\def\unp{\underline{\phantom{a}}\!\!\! p}\def\unP{\underline P}  
\def\unq{\underline{\phantom{a}}\!\!\! q}  
\def\unQ{\underline{\phantom{A}}\!\!\!\! Q}  
\def\unH{\underline{H}}  
  
\def\As {{A \hspace{-6.4pt} \slash}\;}  
\def\bs {{b \hspace{-6.4pt} \slash}\;}  
\def\Ds {{D \hspace{-6.4pt} \slash}\;}
\def\Gts {{\Gt \hspace{-6.4pt} \slash}\;}
\def\ds {{\del \hspace{-6.4pt} \slash}\;}  
\def\ss {{\s \hspace{-6.4pt} \slash}\;}  
\def\ks {{ k \hspace{-6.4pt} \slash}\;}  
\def\ps {{p \hspace{-6.4pt} \slash}\;}   
\def\xs {{x \hspace{-6.4pt} \slash}\;}  
\def\pas {{{p_1} \hspace{-6.4pt} \slash}\;}  
\def\pbs {{{p_2} \hspace{-6.4pt} \slash}\;}   
\def\cFs {{{\cal F} \hspace{-6.4pt} \slash}\;}

\def\Ah{{\hat{A}}}  
\def\Dh{{\hat{D}}}
\def\Gh{{\hat{G}}}
\def\Fh{{\hat{F}}}
\def\Ih{{\hat{I}}} 
\def\Jh{{\hat{J}}} 
\def\Kh{{\hat{K}}}
\def\Lh{{\hat{L}}} 
\def\Ph{{\hat{P}}}
\def\Rh{{\hat{R}}}
\def\Vh{{\hat{V}}} 
\def\Xh{{\hat{X}}}
 
\def\ah{{\hat{\a}}}
\def\bh{{\hat{\b}}}
\def\gh{{\hat{\g}}}
\def\dh{{\hat{\d}}}
\def\hh{\hat{h}}
\def\uh{\hat{u}}  
\def\xh{\hat{x}}  
\def\yh{\hat{y}}  
\def\ph{\hat{p}}  
\def\xih{\hat{\xi}}  
\def\chih{\hat{\chi}}  
\def\Psih{\hat{\Psi}}

\def\psit{\tilde{\psi}}  
\def\Psit{\tilde{\Psi}}   
\def\Psibt{\tilde{\bar{Psi}}}  

\def\st{\tilde{\sigma}}  

\def\delt{\tilde{\delta}}
\def\Phit{\tilde{\Phi}}   
\def\Phitb{\overline{\tilde{Phi}}}  
\def\tht{\tilde{\th}}  
\def\lt{\tilde{\l}}
\def\chit{\tilde{\chi}}   
\def\phit{\tilde{\phi}} 

\def\At{\tilde{A}}
\def\Bt{\tilde{B}}
\def\Ct{\tilde{C}}
\def\Dt{\tilde{D}}
\def\Et{\tilde{E}}
\def\Ft{\tilde{F}}
\def\Gt{\tilde{G}}
\def\Ht{\tilde{H}}
\def\It{\tilde{I}}
\def\Jt{\tilde{J}}
\def\Qt{\tilde{Q}}  
\def\Rt{\tilde{R}}  
\def\Mt{\tilde{M }}  
\def\Nt{\tilde{N}}   
\def\St{\tilde{S}}
\def\Vt{\tilde{V}}
\def\Xt{\tilde{X}} 
\def\at{\tilde{a}}
\def\ct{\tilde{c}}
\def\dt{\tilde{d}}
\def\htt{\tilde{h}} 
\def\ft{\tilde{f}}
\def\gt{\tilde{g}}
\def\pt{\tilde{p}}  
\def\qt{\tilde{q}}  
\def\vt{\tilde{v}}  
\def\nt{\tilde{n}}  
\def\ut{\tilde{u}}  
\def\wt{\tilde{w}}  
\def\zt{\tilde{z}} 
\def\xt{\tilde{x}} 
\def\yt{\tilde{y}} 
\def\Psit{\tilde{\Psi}}
\def\vphit{\tilde{\varphi}}  

\def\eb{\bar{\epsilon}} 
\def\delb{\bar{\partial}}  
\def\thb{\bar{\theta}}
\def\mub{\bar{\mu}}
\def\lamb{\bar{\l}}
\def\psib{\bar{\psi}}
\def\sb{\bar{\sigma}}
\def\xib{\bar{\xi}}
\def\chib{\bar{\chi}}

\def\Psib{\bar{\Psi}}
\def\Phib{\bar{\Phi}}
\def\Lamb{\bar{\Lambda}}
\def\Sb{{\overline \Sigma}}
\def\cb{\bar{c}}
\def\hb{\bar{h}}
\def\qb{\bar{q}}
\def\wb{\bar{w}}
\def\ub{\bar{u}}
\def\zb{{\bar{z}}}
\def\Hb{\bar{H}}
\def\Qb{{\bar Q}}

\def\Ab{{\overline A}} \def\Bb{{\overline B}} \def\Cb{{\overline C}}  
\def\Db{{\overline D}} \def\Eb{{\overline E}} \def\Fb{{\overline F}}  
\def\Gb{{\overline G}} 
\def\Ib{{\overline I}}  
\def\Jb{{\overline J}} \def\Kb{{\overline K}} \def\Lb{{\overline L}}  
\def\Mb{{\overline M}} \def\Nb{{\overline N}} \def\Ob{{\overline O}}  
\def\Pb{{\overline P}}  \def\Rb{{\overline R}}  
 \def\Tb{{\overline T}} \def\Ub{{\overline U}}  
\def\Vb{{\overline V}} \def\Wb{{\overline W}} \def\Xb{{\overline X}}  
\def\Yb{{\overline Y}} \def\Zb{{\overline Z}}  

\def\fb{{\overline f}}
\def\gb{{\overline g}}
\def\mb{{\overline m}}
\def\lb{{\overline l}}
\def\yb{{\overline y}}

\def\ba{{\bf a}} 
\def\bk{{\bf k}}  
\def\bl{{\bf l}}  
\def\bp{{\bf p}}  
\def\bq{{\bf q}}  
\def\br{{\bf r}}
\def\bt{{\bf t}}
\def\bu{{\bf u}}
\def\bv{{\bf v}}
\def\bx{{\bf x}}  
\def\by{{\bf y}}  
\def\bR{{\bf R}}  
\def\bV{{\bf V}}

\def\bone{{\bf 1}}  

\def\va{{\vec a}}
\def\vk{{\vec k}}
\def\vp{{\vec p}}
\def\vq{{\vec q}}
\def\vx{{\vec x}}
\def\vy{{\vec y}}
\def\vu{{\vec u}}
\def\vv{{\vec v}}

\def\vs{{\vec \sigma}}
\def\vtau{{\vec \tau}}

\newcommand{\ov}[1]{\overrightarrow{#1}}

\def\frA{\mathfrak{A}}
\def\frB{\mathfrak{B}}
\def\frC{\mathfrak{C}}
\def\frD{\mathfrak{D}}
\def\frE{\mathfrak{E}}
\def\frF{\mathfrak{F}}
\def\frG{\mathfrak{G}}
\def\frH{\mathfrak{H}}
\def\frM{\mathfrak{M}}
\def\frN{\mathfrak{N}}
\def\frR{\mathfrak{R}}
\def\frW{\mathfrak{W}}

\def\fra{\mathfrak{a}}
\def\frb{\mathfrak{b}}
\def\frf{\mathfrak{f}}
\def\frg{\mathfrak{g}}
\def\frh{\mathfrak{h}}
\def\frl{\mathfrak{l}}
\def\frs{\mathfrak{s}}
\def\fri{\mathfrak{i}}
\def\frj{\mathfrak{j}}

\def\ma{\mathfrak{a}}
\def\mg{\mathfrak{g}}
\def\mh{\mathfrak{h}}
\def\mR{\mathfrak{R}}
\def\mN{\mathfrak{N}}

\def\d{\delta}\def\D{\Delta}\def\ddt{\dot\delta}  
  
\def\pa{\partial} \def\del{\partial}  
\def\xx{\times}  
\def\uno{\mbox{1 \kern-.59em {\rm l}}}    
  
\def\trp{^{\top}}  
\def\inv{^{-1}}  
\def\dag{{^{\dagger}}}  
\def\pr{^{\prime}}  
  
\def\rar{\rightarrow}  
\def\lar{\leftarrow}  
\def\lrar{\leftrightarrow}  
  
\newcommand{\0}{\,\!}      %this is just NOTHING!  
\def\one{1\!\!1\,\,}  
\def\im{\imath}  
\def\jm{\jmath}  
  
\newcommand{\tr}{\mbox{tr}}  
\newcommand{\slsh}[1]{/ \!\!\!\! #1}  
  
\def\vac{|0\rangle}  
\def\lvac{\langle 0|}  
  
\def\hlf{\frac{1}{2}}  
\def\ove#1{\frac{1}{#1}}  

\def\Box{\square}  
\def\CC {\mathbb{C}}
\def\FF {\mathbb{F}}
\def\RR{\mathbb{R}}
\def\NN{\mathbb{N}}  
\def\ZZ{\mathbb{Z}}  
\def\bb#1{{\bf #1}}  
\def\bcomment#1{}  
\def\bfhat#1{{\bf \hat{#1}}}  
\def\VEV#1{\left\langle #1\right\rangle}  

\newcommand{\ex}[1]{{\rm e}^{#1}} \def\ii{{\rm i}}  

\newcommand{\lrbrk}[1]{\left(#1\right)}
\newcommand{\sfrac}[2]{{\textstyle\frac{#1}{#2}}}
 
\def\stw{{\sqrt{2}}}

\def\rf {{\rm f}}
\def\ri {{\rm i}}
\def\rj {{\rm j}}
\def\rk {{\rm k}}
\def\rl {{\rm l}}
\def\rs {{\scriptscriptstyle \rm S}}
\def\rt {{\scriptscriptstyle \rm T}}

\def\rQ {{\scriptscriptstyle \rm \cQ}}
\def\rR {{\scriptscriptstyle \rm \cR}}

\def\cQb{{\cal \Qb}}
\def\cRb{{\cal \Rb}}
\def\cWb{{\cal \Wb}}

\def\fd {{\rm N}}
\def\afd {{\overline{\rm N}}}

\def \II {I\hspace{-.1em}I\hspace{.1em}}
\def \IIA {\mbox{\II A\hspace{.2em}}}
\def \IIB {\mbox{\II B\hspace{.2em}}}
\def \gs {g^s}
\def \ls {\lambda^s}

\def \I {{\cal I}}
\def \qs {q\hspace{-.53em}/\hspace{.15em}}
\def \ks {k\hspace{-.53em}/\hspace{.15em}}
\def \YM {{\mbox{\tiny YM}}}
\def \gym {g_{\YM}}

\def \Lc {\L_c}
\def\IR{\relax{\rm I\kern-.18em R}}
\def \id {{\bf 1}}

\def\cci{\ell}
\def\ccj{\ell'}

\author{Chong-Sun Chu and Gurdeep S. Sehmbi\\  
Centre for Particle Theory
and Department of Mathematical Sciences, 
Durham University, Durham, DH1 3LE, UK \\
E-mail:  
\email{chong-sun.chu@durham.ac.uk}, \email{g.s.sehmbi@durham.ac.uk} }

\title {
Open M2-branes with Flux and Modified Basu-Harvey Equation
}

\abstract{ 
The supersymmetric actions of closed multiple M2 branes with flux 
for the BL and the ABJM theories have  been constructed recently by Lambert
and Richmond in \cite{LR}. In this paper we extend the construction to
the case of open M2-branes with flux and derive the boundary
conditions. %c3 As advocated in \cite{CS1}, this 
This allows us to derive the
modified Basu-Harvey equation in the presence of flux. 
As an example, we consider the Lorentzian BL model. A new feature
of the fuzzy funnel solution describing a D2-D4 intersection 
is obtained as a result of the flux.
}
 
\preprint{DCPT-10/59}
\keywords{M-Theory, D-branes, M-branes}

\begin{document}

\section{Introduction}

The understanding of the physics of branes in M-theory is one of the 
most intriguing and mysterious tasks in string/M theory, see for example 
\cite{berman}.
Recently tremendous
progress has been made in the description of multiple M2-branes 
\cite{BL1,Gut,BL2,BL3,ABJM,BL4}. A major part of the excitement is due to the 
employment of  a novel mathematical structure, the Lie 3-algebra, in the
description of the gauge symmetry of the parallel M2-branes. Although the 
Lie 3-algebra is not essential in the $\cN=6$ description of the M2-branes
\cite{ABJM}, 
%S4 I changed this a few times, plural of evidence is evidence
there is evidence \cite{entropy,CS1,furuuchi,tensor,M5fromM2,HIMS,n-ary} 
that it may have a deeper connection to
M-theory in general.

%S4 based
In a recent paper, based on the earlier works \cite{gomis,hll,gomis2,hll2},
Lambert and Richmond \cite{LR} were able to construct a 
coupling of closed multiple M2-branes 
to specific configurations of background 3-form and
6-form gauge fields of eleven-dimensional supergravity. 
The coupling 
%c4 is a ``non-Abelian'' generalization of the usual Wess-Zumino coupling and 
makes essential use of the underlying 3-algebra structure. 
The flux configuration considered in \cite{LR} 
is self-dual in the space transverse to the
M2-branes and gives rise to a mass term and its supersymmetrization on the
worldvolume theory of the M2-branes. 

In this paper we extend the construction of \cite{LR} to
the open case
and  derive the supersymmetric boundary conditions. As advocated in 
\cite{CS1,CS2}, see also \cite{GW}, 
the boundary condition can be interpreted as an equation of
motion for the boundary fields and hence, in the case of the scalar fields, 
can be understood as describing the non-trivial shape of the boundary of the 
M2-branes. In particular, for a certain specific configuration, the boundary 
condition describes the M2-brane ending on an M5-brane and hence can be 
identified with
the Basu-Harvey equation \cite{BH}. 
We show that this continues to be the case in the 
presence of the flux background.

The plan of the paper is as follows. In section 2, we analyse the 
Bagger-Lambert (BL) action 
coupled to flux and obtain the supersymmetric boundary condition. 
The boundary condition only has
%S4 a
a trivial solution in general. However for
specific configurations of the scalar fields 
we consider, the boundary condition is non-trivial.
For example, one obtains a mass deformed Basu-Harvey equation
which describes a 
system of M2-branes ending on an M5-brane in the presence of a background flux.
The analysis of the boundary condition also asserts the absence of 
supersymmetric M2-M9 intersection in the  presence of flux.
In section 3,
we perform the same analysis for the ABJM theory with flux. In section 4, we 
consider the flux modified BL
theory with Lorentzian 3-algebras. 
In the closed case, the 
theory is equivalent to the $\cN=8$ supersymmetric Yang-Mills theory; thanks
to the complete decoupling of one of the scalar fields, say
$X^{10}$, of the eight scalar fields in the BL theory. This is no 
longer automatic
in the open case and a boundary condition has to be chosen to achieve this. 
The resulting theory then describes  multiple D2-branes in a 
mixed NS-NS and R-R flux background. The system is however not supersymmetric
in general since the decoupling boundary condition generally breaks  
the supersymmetry. We 
%c4 then 
consider  a particular configuration of the scalar
fields and show that a supersymmetric boundary condition can be obtained. 
We show that this describes a system of D2-branes ending on a D4-brane in the 
the background flux. New features of the fuzzy funnel solution are 
also discussed.

\section{Boundary Condition for the BL Theory Coupled to Flux}

\subsection{$\cN=8$ closed M2-branes in flux background}

In this subsection, we review the construction 
%S4 removed of - used too many times.
 \cite{LR} of 
the supersymmetric action which describes the 
coupling of multiple closed M2-branes to a certain configuration of 
background gauge field.
In the limit of large $T_{M2}$,
the Lagrangian consists of a flux and 
a mass term modification to the 
Bagger-Lambert Lagrangian $\cL_{\cN=8}$:
\be \label{LR1}
\cL = \cL_{\cN=8} + \cL_{flux} + \cL_{mass},
\ee
where 
\bea 
\cL_{\cN=8} &=& -\frac{1}{2} \Tr(D^\m X^{I}, D_\m X^I) 
+ \frac{i}{2}\Tr( \Psib, \G^\m D_\m \Psi) 
+ \frac{i}{4} \Tr( \Psib, \G_{IJ} [X^I, X^J, \Psi]) \nn\\
&& -\frac{1}{12} \Tr \big( [X^I,X^J,X^K],  [X^I,X^J,X^K]\big) + \cL_{CS}, 
\label{L-BL}\\
\cL_{flux} &=& 
c \Gt_{IJKL} \Tr(X^I, [X^J,X^K,X^L]) , \label{L-flux} \\
\cL_{mass} &=& -\frac{1}{2} m^2 \delta_{IJ}\Tr(X^I,X^J) 
+ b\Tr(\Psib\G^{IJKL}, \Psi)\tilde{G}_{IJKL} \label{L-mass}
\eea
and $\Tr(\cdot,\cdot)$ is the metric for the Lie 3-algebra. 
The background gauge field has the transverse components $G_{IJKL}$ 
turned on and  
$\Gt_{IJKL}$ is defined by
\be 
\Gt_{IJKL} 
% = -\frac{1}{3!}\e^{\m\n\l} ( G_7 + \frac{1}{2} C_3 \wedge G_4)_{\m\n\l IJKL} 
= \frac{1}{4!}\e_{IJKLMNPQ} G^{MNPQ},
\ee
where $I,J,K,L =3,4,\cdots, 10$.
%c3 change name of susys so as to be consistent with lorentzian 3-alg case
The supersymmetry transformation is given by $\d =\d_0 + \d'$ where
$\d_0$ is the supersymmetry transformation of the original $\cN=8$ theory,
\bea
\d_0 X^I_a &=& i \eb \G^I \Psi_a, \label{d1}\\
%S4 put empty bracers to position the index a on the f.
\d_0 \At_\m{}^b{}_a &=&  i \eb \G_\m \G_I X^I_c \Psi_d f^{cdb}{}_a, \label{d2}\\
\d_0 \Psi_a &=& D_\m X^I_a \G^\m \G^I \e 
- \frac{1}{6} X^I_b X^J_c X^K_d f^{bcd}{}_a \G^{IJK} \e, \label{d3}
\eea
and $\d'$ is the  additional contribution to the supersymmetry 
transformations due to the flux
\bea
\d' X^I_a &=& 0,  \label{dp1} \\
\d' \At_\m{}^b{}_a &=& 0, \label{dp2}\\
\d' \Psi_a &=& \omega \G^{IJKL}\G^M \e X_a^M \tilde{G}_{IJKL}. \label{dp3}
\eea
Here 
$\Psi$ and $\e$ are eleven dimensional spinors satisfying the conditions
\be \label{c0}
\G_{012} \Psi =- \Psi, 
\ee
\be \label{c0p}
\G_{012} \e= \e.
\ee 
Inclusion of the effects of the backreaction of the flux implies that 
$c=2$ \cite{LR}. 
Supersymmetry requires  
the coefficients $\o$ and $b$ to be  determined by the 
flux term 
\be \label{c2} 
\o = \frac{c}{8}, \quad b = -i \frac{c}{16}. 
\ee
Moreover, the flux $\Gt_{IJKL}$
has to be  self-dual, which implies that
\be \label{G1}
\G^{012} \Gts = \Gts,
\ee
where $\Gts \equiv \G^{IJKL} \Gt_{IJKL}$. 
It also needs to satisfy the condition
\be\label{G2}
\Gts \Gts =  8m^2 (1+ \G^{3456789(10)} ),
\ee
which implies immediately that
\be \label{c1}
G_{MN[IJ}G_{KL]}{}^{MN} =0
\quad \mbox{and}
\quad
m^2 = \frac{c^2}{32 \cdot 4!} G^2.
\ee

The self-duality condition is solved by $\Gts$ of the form
$ \Gts = d (1+ \G^{012}) R$, 
where $d$ is a constant coefficient and $R$ is a sum of products of four
transverse $\G^I$'s, $I= 3, 4, \cdots, 10$. The condition \eq{G2} then implies 
that
\be \label{gen-G}
\Gts = 2 \mu \frac{1+ \G^{012}}{2} R, \quad R^2 =1,
\ee
%c4
for $\mu = \pm 2 m$, $m\geq 0$.
A simple solution is 
\be
R = \G^{3456}. 
\ee
This corresponds to  the flux
\be
\Gt =  \mu \;dx^3\wedge dx^4\wedge dx^5\wedge dx^6 + \mbox{dual},
\label{sf1}
\ee
and the Lagrangian \eq{LR1} reproduces precisely the deformed Bagger-Lambert 
Lagrangian of \cite{gomis} and \cite{hll}.

\subsection{Flux modified supersymmetric boundary 
condition}

Next we want to consider the open case of the 
flux modified BL theory and derive the boundary condition.
Note that in the above derivation of the supersymmetric 
invariance of the action,
boundary contributions have been dropped due to the 
%c4 closed conditions 
closedness of the M2-branes. 
These boundary terms have to be kept carefully in the
presence of a boundary.
It is easy to see that these contributions arise from the fermion and scalar
kinetic terms in the Lagrangian $\cL_{\cN=8}$. We have
\be
\d \cL = \frac{i}{2}\del_{\mu}\Tr(\Psib\G^\m,\d \Psi) 
- \del_\m \Tr(\d X^I, D^\m X^I  ) + \mbox{bulk terms}, 
\ee
where the ``bulk terms'' denote non-total derivative terms and are precisely
equal to zero when the conditions 
%c4 
\eq{c2}--\eq{c1} are satisfied. 
To proceed, let 
us consider the M2-branes to have a boundary at $\s_2 =0$. 
%c2
We have 
\be
\d \int d^3 \s \cL = \frac{i}{2}\int d^2 \s\left(
\Tr(\Psib\G^{2},\d\Psi) 
-2\Tr(D_{2}X^I, \Psib\G^I\e) \right).
\ee
%S4 I dont see why we need to use this? Using the condition \eq{c0}, 
We obtain 
the boundary condition
\be
\label{beom}
 0=D_{\a}X^I\Psib\G^2\G^\a \G^I\e 
-\frac{1}{6}[X^I,X^J,X^K]\Psib\G^2\G^{IJK}\e 
+\frac{1}{4} X^M\Psib\G^2\Gts\G^M\e  -D_2X^I\Psib\G^I\e, 
\ee
where $\a = 0,1$ and 
%c2 we carefully drop the $\Tr(\cdot,\cdot)$ notation. 
the trace $\Tr$ is understood. 
This is the most general supersymmetric 
boundary condition one may have for a system of open M2-branes in 
our flux background. In general, due to the different number of $\G$-matrices
in each term, the equation \eq{beom} generically only has trivial solution.  
Non-trivial solutions can be obtained only when 
%c4 
additional conditions are imposed on the matter fields and on the supersymmetry
parameters.

The analysis of the boundary condition in the absence of flux was performed 
in \cite{BermanBH}, where the solutions to the boundary condition
are classified according to the number of scalars obeying a Dirichlet condition
(or more precisely being set equal to zero). In the following 
we perform a similar 
analysis for the boundary condition \eq{beom} with flux
and determine what 1/2 BPS 
configurations are allowed as endpoint of the system of open
M2-branes. To be specific, we
will consider the flux configuration \eq{sf1}. 

\subsubsection{Half Dirichlet: flux modified Basu-Harvey equation}
 
This case corresponds to  
an ansatz where half of the scalar fields are set to zero, for 
example, 
\be \label{r1}
X^{3,4,5,6} = 0.
\ee
This means that we have reduced the $SO(8)$ to an $SO(4)$ 
R-symmetry. Let us also impose the projection condition
\begin{equation}
\label{m5}
 \G^{01789(10)}\e = \e.
\end{equation}
It follows immediately that
\be
%S4 changed the 2 and the l around as discussed in email
\G^{ijk} \e =\ve^{ijkl}\G^2\G^l\e, \quad i,j,k,l = 7,8,9.10.
\ee
We also have 
\be
\G^{2}\Gts\e = 2 \mu  \, \e.
\ee
%c4 It follows that the 
The boundary condition \eq{beom} is then reduced  to
\be
\label{preind}
 0 =D_{\a}X^i\Psib\G^2\G^\a\G^i\e 
 %S4 minus sign as a result of changed relation above
-\frac{1}{6}\e^{ijkl}\Psib\G^l\e[X^i,X^j,X^k] 
+\frac{\m}{2} \Psib\G^i\e X^i 
- D_2X^i\Psib\G^i\e.
\ee
We note that the first term in \eq{preind} is identically zero 
if we also impose the condition on the fermion
\be
\label{ferm5}
\G^{01789(10)}\Psi = -\Psi.
\ee
This reduction in the degrees of freedom is compatible with the 1/2 BPS nature 
of the projector \eq{m5}. 
As a result, we obtain the boundary equation of motion
\be
\label{r2}
 D_2X^i = -\frac{1}{6}\varepsilon^{ijkl}[X^j,X^k,X^l] +  \frac{\m}{2} X^i
\ee
for $ i,j,k,l = 7,8,9,10$.
This is the Basu-Harvey equation modified by the flux \eq{sf1}.

We must now check that the boundary conditions \eq{r1}, \eq{ferm5} and \eq{r2}
%c4 are invariant under the supersymmetry transformations. 
are supersymetric invariant.
Indeed it is easy to verify that
\be
\delt X^{i'} =0, \quad i' =3,4,5,6 
\ee
and 
\be
(1+ \G^{01789(10)})\; \delt \Psi =0
\ee
using the conditions \eq{m5} and \eq{ferm5}. 
As for \eq{r2}, supersymmetry requires the fermionic boundary equation
\be \label{r3}
D_2 \Psi + \frac{1}{2}\Gamma^2 \Gamma_{ij}[\Psi,X^i,X^j] -\frac{\m}{2}\Psi =0.
\ee

Note that if instead of \eq{m5}, we preserve the other half of supersymmetry 
\be \label{m5p}
 \G^{01789(10)}\e = - \e,
\ee
then exactly the same analysis results in the other Basu-Harvey equation.
\be \label{r2p}
D_2X^i = \frac{1}{6}\varepsilon^{ijkl}[X^j,X^k,X^l] -  \frac{\m}{2} X^i.
\ee
This is equivalent to  \eq{r2} with $x^2 \to -x^2$. 

The modified Basu-Harvey equation can also be derived as
the  Bogomoln'yi bound for the system of closed M2-branes \cite{BL3}. 
In fact,  
consider static solutions that depend on one coordinate
$\s_2=s$, the energy can be written as
\be \label{BL-E}
E = \frac{1}{2} \int d \s_1 d s\;
\Big[
\Tr \big(
\frac{d X^i}{ds} \pm \del^i W, \frac{d X^i}{ds} \pm \del^i W 
\big)
\mp \Tr \big( 2 \del^i W, \frac{d X^i}{ds}\big)  
\Big],
\ee
%c4  W corrected. 
where  $W= -\frac{\mu}{4} (X^i)^2 + \frac{1}{24}\ve^{ijkl}\Tr(X^i,[X^j,X^k,X^l])$
and $\Tr(T_a , T_b) = g_{ab}$ is the metric of the Lie 3-algebra. 
The ``+'' choice in the first term on \eq{BL-E} gives 
the modified Basu-Harvey equation \eq{r2}, while the ``-'' choice 
gives the other BPS equation \eq{r2p}.
Once again 
we have seen the power of 
utilizing the boundary system. For other applications of using the 
boundary system see, for example, \cite{CS1,CS2,BermanBH}.

The condition \eq{r2} (or \eq{r3}) represents a non-trivial 
boundary condition of the system of open M2-branes. For an $\cA_4$ 
Lie 3-algebra, 
%c4 $[T^i,T^j,T^k] = \ve^{ijkl} T^l$, using the ansatz $X^i(s) = f(s) T^i$, 
% one obtains
% \be \label{f-de} f'- f^3 - \frac{\mu}{2} f=0. \ee
two kinds 
of solution were found \cite{BL3}. 
One describes a domain wall interpolating
between two 
%S4 plural for vacuum
vacua of the worldvolume theory, the other is a fuzzy funnel 
solution that describes a system of M2-branes ending on a single M5-brane.
For 
%S4 a
a more general Lie 3-algebra, the Basu-Harvey equation \eq{r2} still 
%S4 admits
admits these
solutions if $\cA_4$ can be found as a subalgebra. For 
%S4 a
a Lie algebra, 
semisimplicity guarantees that a Lie algebra always 
%S4 admits
admits an $su(2)$ subalgebra.
 The question of when a 
Lie 3-algebra has an $\cA_4$ subalgebra is an interesting and open one 
\cite{cw}.

\subsubsection{No Dirichlet: absence of supersymmetric M9-brane}

In this case we keep all the eight scalar fields and do not
assume 
%S4
%any of them to be zero. 
that any of them are zero.
Consider 
%S4 Imposing
imposing the projection condition
\be
\G^{013456789(10)}\e = \e,
\ee
which implies
\be
%S4 added \e at the end
\G^2\e = \e = \G^{01}\e.
\ee
%c4 no need to do this
% We note the following relation using the above to simplify the boundary 
% equation of motion
% \be
% \G^{IJK}\e = -\varepsilon^{IJKLMNOP}\G^{LMNOP}\e,
% \ee
% where $I,J,K,L,M,N,O,P = 3,...,10.$
% We may now use the above in the boundary equation of motion for 
% open M2-branes \eq{beom} to obtain
Also we impose the condition
\be
\G^{013456789(10)}\Psi = \Psi
\ee
on the fermion. 
The first term in \eq{beom} is again zero using these conditions.
The 
%S4 did you mean second to last?
second to last term in \eq{beom} is a linear combination of 
%S4 products
products of three or 
five transverse Gamma matrices $\Gamma^I$. 
One can see that the equation \eq{beom} contains
different number of transverse $\Gamma^I$'s and linear 
independence of them implies that
\be
%S4 rm .
D_2 X^I = 0
\ee
and
\be \label{xzero-all}
X^I =0.
\ee
%S4 capitalize There here?
i.e. There is no nontrivial solution.

Notice that if we turn off the flux ($m=0$), then the 
boundary equation \eq{beom}
can be solved 
%S4 hyphen
non-trivially with 
\be
D_2 X^I = 0, \quad [X^I,X^J,X^K] =0.
\ee
This has been  interpreted as
%S4
an M9-brane occupying the directions
013456789(10) where the M2-branes end on\cite{BermanBH}.

With the  flux \eq{sf1} turned on, however, we only get the 
trivial solution \eq{xzero-all}. This means, 
%c4 
in the presence of flux,
the system of M2-branes cannot end on an M9 brane supersymmetrically.
This is a prediction of our open M2-branes analysis. 
One way to confirm its 
validity is 
to determine the supersymmetry projector of an M9-brane in 
the presence of flux and show that the preserved supersymmetry is 
incompatible with the M2-brane supersymmetry projector $\G^{012} \e =\e$.
To carry out this analysis, one needs 
%c4 
to first construct 
the supergravity solution of M9-brane with a constant flux 
and then  determine the 
preserved supersymmetry as performed in \cite{berg} for the case
without flux. 
%c5
One can also reduce the system down to 10 dimensions on $x^{10}$. This becomes
a D2-D8 intersection. 
%S4 endowed
The D8-brane is endowed with a worldvolume NS-NS $B$-field
in the 78, 79 or 89 directions and the preserved 
supersymmetry is determined by
%S4 Added 1 direction
\be
e^{-a/2} \G^{013456789(10)}e^{a/2} \e =\e, 
\ee
where $a = \frac{1}{2}Y_{IJ}\G^{IJ} \G^{(10)}$
and $Y$ is a nonlinear fumction of $B$ whose explicit form can be found in
\cite{d1}. What is important to us is that only the 78, 79 or 89 components are 
nonzero in our case.
It is then clear that the supersymmetry perserved by the D8-brane
is incompatible with
%S4 removed 'that'
 $\G^{012} \e =\e$ of the D2-brane.
Therefore the D2-D8 system and the M2-M9 system are not supersymmetric. 

\subsubsection{All Dirichlet: M-wave}

In this case we set all the eight scalars to zero at the boundary. 
As a result, all the modifications due to flux vanish and 
the boundary conditions read identically as in the flux-less case
\be
D_2 X^I \Psib \G^I \e=0.
\ee
This can be solved immediately if one imposes the projection conditions
\be
(1-\G^2) \e =0, \quad (1+ \G^2) \Psi =0.
\ee
The solution has been interpreted as an 
M-wave where the M2-branes end on \cite{BermanBH}.

\section{Boundary Condition for the ABJM Theory Coupled to Flux}

\subsection{$\cN=6$ closed M2-branes in flux background}

We now turn to the $\cN=6$ theory with mass and flux terms 
given by Lambert-Richmond \cite{LR} and discuss the boundary 
terms and their implications. The full Lagrangian of the flux 
deformed $\cN=6$ theory reads 
\be
\label{LRN6}
\cL = \cL_{\cN=6} + \cL_{flux} + \cL_{mass}.
\ee
where
\bea 
\cL_{\cN=6} &=& -\Tr( D^\m \bar{Z}_A ,D_\m Z^A) 
-i\Tr( \psib^A, \g^\m D_\m \psi_A )
-V +\cL_{\mathrm{CS}} \nn\\
&& -i\Tr (\psib^A,[\psi_A,Z^B;\bar{Z}_B])
+2i\Tr (\psib^A,[\psi_B,Z^B;\bar{Z}_A]) \nn\\
&& +\frac{i}{2}\,\ve_{ABCD}\Tr(\psib^A,[Z^C,Z^D;\psi^B]) 
-\frac{i}{2}\,\ve^{ABCD}\Tr(\bar{Z}_D,[\psib_A,\psi_B;\bar{Z}_C]), 
\label{L6-BL}\\
\cL_{\mathrm{CS}} &=& \frac{k}{4 \pi} \e^{\m\n\l} \left(
A_\m \del_\n A_\l + \frac{2i}{3} A_\m A_\n A_\l
- 
\Ah_\m \del_\n \Ah_\l - \frac{2i}{3} \Ah_\m \Ah_\n \Ah_\l
\right),\label{L6-CS} \\
\cL_{flux} &=& 
\frac{c}{4} \Tr(\bar{Z}_D,[Z^A,Z^B;\bar{Z}_C]) \Gt_{AB}{}^{CD} , 
\label{L6-flux} \\
\cL_{mass} &=& -m^2 \Tr(\bar{Z}_A,Z^A) 
+ b\Tr(\psib^A, \psi_F)\Gt_{AE}{}^{EF}. \label{L6-mass}
\eea
The $\cN=6$ theory has a 3-algebra given by a matrix representation
\be
[Z^A,Z^B,\Zb_C] = \frac{2\p}{k}(Z^A\bar{Z}_C Z^B - Z^B\bar{Z}_C Z^A),
\ee
which is only antisymmetric in the first two indices.
Here $V$ is defined in \cite{BL4} and we also define 
\be
\Gt_{AB}{}^{CD} = \frac{1}{4}\ve_{ABEF}\ve^{CDGH}G^{EF}{}_{GH}.
\ee
and $A,B,C,D = 1,..,4$ are the $SU(4)$ R-symmetry indices. 
The supersymmetry transformations of the original $\cN=6$ theory is given by
%c3 change susy notation
\bea
\d_0 Z^A &=& i \eb^{AB}\psi_{B}, \label{6d1}\\
\d_0 A_\m &=&  \frac{2\pi}{k}\left[Z^B\psib^A\g_\m\e_{AB} 
+ \e^{AB}\g_\m\psi_A\bar{Z}_B\right], \label{6d2}\\
\d_0 \psi_{A} &=& \g^\m\e_{AB}D_\m Z^B +N_A, \label{6d3}
\eea
and their conjugates, where 
\be
N_A = \frac{2\pi}{k}\left[-\e_{AB}\left(Z^C\bar{Z}_CZ^B 
-Z^B\bar{Z}_CZ^C \right) +2\e_{CD}Z^c\bar{Z}_AZ^D\right]
\ee
and $\d'$ is the  additional contribution to the supersymmetry 
transformations due to the flux
\bea
\d' Z^A &=& 0,  \label{6dp1} \\
\d' A_\m &=& 0, \quad \d' \Ah_\m =0, \label{6dp2}\\
\d' \psi_{A} &=& \omega \e_{DF} Z^F \Gt_{AE}{}^{ED}. \label{6dp3}
\eea
The symmetry transformation parameter satisfies the reality condition
\be
\e_{FP} = \frac{1}{2} \ve_{IJFP} \e^{IJ}.
\ee
For the action to be supersymmetric, the flux needs to take the form 
\be
\label{n6flux}
\Gt_{AB}{}^{CD} = \frac{1}{2}\delta^C_B\Gt_{AE}{}^{ED}
 -\frac{1}{2}\delta^C_A\Gt_{BE}{}^{ED} 
-\frac{1}{2}\delta^D_B\Gt_{AE}{}^{EC} + \frac{1}{2}\delta^D_A\Gt_{BE}{}^{EC},
\ee
where the matrix $\Gt_{AE}{}^{EB}$ has to be traceless
 $\Gt_{AE}{}^{EA} =0$ 
and squares to one
\be
\Gt_{AE}{}^{EB} \Gt_{BF}{}^{FC} = \frac{m^2}{\o^2} \d^C_A.
\ee
Supersymmetry also relates the coefficients $\o, b, m$ to the flux term:  
\be
\o = \frac{c}{4}, \quad b = -i \frac{c}{4}, 
\quad m^2 = \frac{c^2}{32 \cdot 4!} G^2,
\ee
where $G^2 = 6 G_{AB}{}^{CD} G^{AB}{}_{CD}$. As before, one finds
$c=2$ by a  backreaction analysis.

Taking the flux  
\be \label{Gform}
\Gt_{AE}{}^{ED} = 
\begin{pmatrix}
  \ \m & \ 0 & 0 & 0 \\
  0 & \m & 0 & 0 \\
  0 & 0 & -\m & 0 \\
  0 & 0 & 0 & -\m
\end{pmatrix},
\ee
%c4
for $\mu = \pm 2m$, $m\geq 0$,  
one obtains immediately the deformed theory in \cite{gomis2,hll2}. 

\subsection{
Flux modified Basu-Harvey equation}

We now proceed by finding the boundary contributions to 
the $\cN=6$ theory of open M2-branes probing the orbifold 
$\mathbb{C}^4/\mathbb{Z}_k$. 
Again  the contributions for the 
boundary come from total derivative terms in the ABJM theory, 
these arise from the scalar and fermionic terms once again. So we have 
\be
\d \cL = -2\del_\m \Tr(\d\bar{Z}_A,D^\m Z^A) 
- i\del_\m \Tr (\psib^A,\g^\m \d\psi_A) + \mbox{bulk terms},
\ee
Imposing the boundary condition $\s_2=0$ yields the 
boundary equations of motion
\bea
\label{bvn6}
0=&-&2i\Tr \Big(\psib^B \e_{AB},D^2 Z^A\big) 
- i\Tr\Big[\psib^A\g^2, \g^\m \e_{AB}D_\m Z^B \nn\\ 
&+& \frac{2\pi}{k}\Big(-\e_{AB}(Z^C \bar{Z}_C Z^B  - Z^B \bar{Z}_C Z^C )   
+ 2\e_{CD}Z^C\bar{Z}_A Z^D\Big) + \omega \e_{DF} Z^F \Gt_{AE}{}^{ED} \Big].
\quad \qquad
\eea
Now the flux \eq{Gform} can be written compactly as 
\be
\Gt_{AE}{}^{ED} = \m \d_A^D\eta_A,
\ee
where $\eta_A$ is a sign defined as 
\be
  \eta_A = 
  \begin{cases}
   +1 & \text{if $A=1,2$} \\
   -1 & \text{if $A=3,4$}
  \end{cases} .
\ee
Using this in \eq{bvn6}, we obtain the boundary equation of motion
%c4 restore \mu, Psi -> psi, restore missing \g^2
\bea \label{bcc1}
0= \psib^A\Big[&&2D^2 Z^B -\g^2 \g^\m D_\m Z^B 
+ \g^2 \frac{2\pi}{k} \left(Z^C \bar{Z}_C Z^B  
- Z^B \bar{Z}_C Z^C \right)  \nn \\ &&- \frac{\m}{2} \, \eta_A\g^2 Z^B \Big] 
\e_{AB}
-\frac{4\pi}{k}\psib^F\g^2 \e_{AB} Z^A \bar{Z}_F Z^B,
\eea
where we have now suppressed the traces and will imply them 
%S4 implicitly 
in the natural way henceforth. 
This is the most general 
%c4
supersymmetric
boundary equation of motion for 
open M2-branes in the $\cN=6$ theory with our specific flux configuration.

To analyse the boundary condition, it is convenient to
introduce the following notation $A=(a,i)$ and 
denote $Z^A=(X^a,Y^i),\ \psi_A = (\chi_a,\xi_i)$, where $a=1,2$  corresponds 
to the directions 3456 and $i=1,2$ corresponds to the directions 789(10).  
The supersymmetry parameter $\e_{AB}$ is in the $\textbf{6}$ 
representation of $SU(4)$, and it decomposes as \cite{BermanBH}
\be
\e_{AB} = 
\begin{pmatrix}
  \ \varepsilon_{ab}\e & \e_{ai} \\
  -\e_{ai} & \ \varepsilon_{ij}\tilde{\e}
\end{pmatrix}.
\ee
Using the new notation, the boundary condition \eq{bcc1} 
splits into four equations
\bea\label{1of4}
0= \chib^a\varepsilon_{ab}\Big[&&2D^2 X^b -\g^2 \g^\m D_\m X^b 
+ \g^2 \frac{2\pi}{k} \left(Z^C \bar{Z}_C X^b  - X^b \bar{Z}_C Z^C \right) 
- \frac{\m}{2} \g^2 X^b \Big] \e \nn  \\
&&-\frac{4\pi}{k}\e_{ab}\psib^F\g^2\e X^a \bar{Z}_F X^b, 
\eea
\bea \label{2of4}
0= \chib^a\Big[&&2D^2 Y^i -\g^2 \g^\m D_\m Y^i 
+ \g^2 \frac{2\pi}{k} \left(Z^C \bar{Z}_C Y^i  - Y^i \bar{Z}_C Z^C \right) 
- \frac{\m}{2} \g^2 Y^i \Big] \e_{ai} \nn \\
&&-\frac{4\pi}{k}\psib^F\g^2\e_{ai} X^a \bar{Z}_F Y^i, 
\eea
\bea \label{3of4}
0=\xib^i\Big[&&2D^2 X^a -\g^2 \g^\m D_\m X^a 
+ \g^2 \frac{2\pi}{k} \left(Z^C \bar{Z}_C X^a  - X^a \bar{Z}_C Z^C \right) 
+ \frac{\m}{2} \g^2 X^a \Big] \e_{ai} \nn \\
&&-\frac{4\pi}{k}\psib^F\g^2\e_{ai} Y^i \bar{Z}_F X^a,
\eea
\bea
\label{4of4}
0= \xib^i\varepsilon_{ij}\Big[&&2D^2 Y^j -\g^2 \g^\m D_\m Y^j 
+ \g^2 \frac{2\pi}{k} \left(Z^C \bar{Z}_C Y^j  - Y^j \bar{Z}_C Z^C \right) 
+ \frac{\m}{2} \g^2 Y^j \Big] \tilde\e \nn \\
&&-\frac{4\pi}{k}\e_{ij}\psib^F\g^2\tilde\e Y^i \bar{Z}_F Y^j.
\eea

%c4
Let us consider the half Dirichlet case by setting half of the scalars zero,
in particular, $Y^i=0$. 
%S SU(4) -> SU(2) R-symm
This condition reduces the R-symmetry from $SU(4)$ to $SU(2)$.
We first analyze \eq{1of4} and \eq{3of4}. It turns out that  
the second term  ``$\g^\m D_\m X^b$''  in these equations
vanishes for $\mu =0,1$. We will come back to this later. For the moment, 
assuming that this is true, then \eq{1of4} and \eq{3of4} becomes
\be
\label{equiv1}
0= \chib^a\varepsilon_{ab}\Big[D^2 X^b 
+ \g^2 \frac{2\pi}{k} \left(X^c \bar{X}_c X^b  - X^b \bar{X}_c X^c \right) 
- \frac{\m}{2}\g^2 X^b \Big] \e 
-\frac{4\pi}{k}\ve_{cd}\chib^a \g^2 \e X^c \bar{X}_a X^d
\ee
and
\bea
\label{equiv2}
0=\xib^i\Big[&&D^2 X^a + \g^2 \frac{2\pi}{k} \left(X^c \bar{X}_c X^a  
- X^a \bar{X}_c X^c \right) + \frac{\m}{2}\, \g^2 X^a \Big] \e_{ai}.
\eea
The two equations are not compatible with each other in general. However
it is possible to impose a suitable  supersymmetry projection 
conditions on the spinors $\e$ and $\e_{ai}$ so that these two equations become
equivalent. The needed conditions are
\bea 
&&(1+\g^2)\e = 0 = (1+\g^2)\tilde\e \label{proj1}\\
&&(1-\g^2)\e_{ai} = 0, \label{proj2}
\eea
or 
\bea 
&&(1-\g^2)\e = 0 = (1-\g^2)\tilde\e \label{proj1p}\\
&&(1+\g^2)\e_{ai} = 0, \label{proj2p}
\eea
%c4 It is simple to show that 
As a result, \eq{equiv1} and \eq{equiv2} are identical since
\be
X^c \bar{X}_c X^b  - X^b \bar{X}_c X^c = 
\varepsilon^{ba}\varepsilon^{cd}X^c\bar{X}_a X^d.
\ee
And we obtain the modified Basu-Harvey equation with 
mass,
\be \label{abjm-BH}
D_2 X^a \pm \frac{2\pi}{k}\left(X^c \bar{X}_c X^a  
- X^a \bar{X}_c X^c\right) \pm \frac{\m}{2} X^a=0,
\ee
where the $+$ sign corresponds to the choice \eq{proj1}, \eq{proj2} and
the $-$ sign corresponds to the choice \eq{proj1p}, \eq{proj2p}.
This can also be written in terms of the 3-bracket
\be
\label{n6BH}
D_2 X^a \pm \left[X^c, X^a; \bar{X}_c\right] \pm \frac{\m}{2} X^a=0.
\ee
This is the mass deformed Basu-Harvey equation for the flux modified 
ABJM theory.
In the following, we consider the choice of 
projectors \eq{proj1}, \eq{proj2} and
the Basu-Harvey equation
\be \label{abjm-BHp}
D_2 X^a + \frac{2\pi}{k}\left(X^c \bar{X}_c X^a  
- X^a \bar{X}_c X^c\right) + \frac{\m}{2} X^a=0.
\ee 
The
analysis for the other choice is exactly the same.
 
Next we note that since 
$\d Y^i = i \e^{ia} \chi_a + i \ve^{ij} \tilde \e \xi_j$, the boundary 
condition $Y^i =0$ is supersymmetric invariant, 
%c4 add commas
after imposing \eq{proj1} and \eq{proj2}, if 
\bea
(1+\g^2)\chi_a &=&0, \label{proj3}\\
(1-\g^2)\xi_i &=&0. \label{proj4}
\eea
As for the conditions \eq{2of4} and \eq{4of4}, which read
\be
\chib^a \e_{ai} D_2 Y^i =0,
\ee
and
\be
\xib^i \ve_{ij} \tilde\e D_2 Y^j =0.
\ee
These are satisfied immediately as a result of the projection conditions
\eq{proj1}, \eq{proj2}, \eq{proj3}, \eq{proj4}.
It is also easy to see that these projection conditions 
are supersymmetric invariant.
Moreover, supersymmetry on \eq{abjm-BH} requires the fermionic boundary 
equations
\be
D_2 \chi_c - [\chi_c,X^a;\bar{X}_a] + 2[\chi_d,X^d;\bar{X}_c]
-\frac{\m}{2} \chi_c =0,
\ee
\be
D_2 \xi_j + [\xi_j,X^a;\bar{X}_a] - \ve_{jk} \ve_{ab} [X^a,X^b;\xi_k]
-\frac{\m}{2} \xi_j =0.
\ee

%c4 
% \be
% D_2 \chi_c + \g^2[\chi_c,X^a;\bar{X}_a] - 2\g^2[\chi_d,X^d;\bar{X}_c]
% +\frac{\m}{2} \g^2 \chi_c =0,
% \ee
% \be
% D_2 \xi_j + \g^2[\xi_j,X^a;\bar{X}_a] - \ve_{jk} \ve_{ab}\g^2 [X^a,X^b;\xi_k]
% -\frac{\m}{2} \g^2 \xi_j =0.
% \ee

As for the above assumption of the vanishing of the
terms of the form  ``$\G^\m D_\m X^b$''  in the equations
\eq{1of4} and \eq{3of4}, one can see that it follows immediately from
the projection conditions \eq{proj1} and \eq{proj3}, and respectively
\eq{proj2} and \eq{proj4}. 

The Basu-Harvey equation \eq{abjm-BHp} can be readily solved by employing 
the ansatz
\be
X^a(s) = f(s) R^a,
\ee
where $s=x_2$ and $R^a$ are $N\times N$ matrices satisfying the relation
\be \label{Q}
R^c R^\dag_c R^a -R^a R^\dag_c R^c = -R^a.
\ee
Then we obtain
\be \label{f-dep}
f' - \frac{2\pi}{k} f^3 + \frac{\mu}{2} f=0.
\ee
The equation \eq{Q} has been solved in \cite{gomis2} and 
the irreducible solution is
\be
(R^1)_{mn} = \d_{m,n} \sqrt{m-1}, \quad
(R^2)_{mn} = \d_{m-1,n} \sqrt{N-m+1}, \quad m,n =1, \cdots, N.
\ee
A direct sum of such blocks is also a solution. 
The equation \eq{f-dep} is the same equation as in the $\cN=8$
theory. This is how the 
domain wall solution 
and the M2-M5 intersection are represented in the $\cN=6$  theory.

Finally, let us comment briefly on the no Dirichlet and all Dirichlet cases. 
For the no Dirichlet case, we find only 
the trivial solution $X^a=Y^i=0$ as in the $\cN=8$ theory.
As for the  all Dirichlet case,
since  the flux modifications all go away when
all the scalars are set to zero at the boundary, hence the boundary conditions 
\eq{1of4}-\eq{4of4} reduce to exactly the same form as 
in flux-less case  and one 
%S4 gets
gets an M-wave \cite{BermanBH}.

\section{Lorentzian 3-Algebras and a Reduction to D2-Branes}

In the original construction of the BL theory \cite{BL1,BL2,BL3},
the Lie 3-algebra  $\cA_4$ was employed.
The use of 
$\cA_4$  was  motivated by the studies of 
Basu and Harvey \cite{BH} whose main objective was to construct a
generalization of the Nahm equation for describing intersecting M-branes.
The next simplest example of a Lie 3-algebra is the Lorentzian algebra.
It has been shown that when one considers  
a Lorentzian 3-algebra,
the Bagger-Lambert Lagrangian reduces to the 
$\cN=8$ SYM theory of multiple D2-branes
\cite{gomis-lor,v-lor,Ho1,gomis3,schwarz,gv,mukhi},
%c4
as opposed to the nontrivial reduction for the original BL theory \cite{MP}. 
In this section, we consider the
Lorentzian BL theory with flux and analyse its reduction. We will 
also derive the supersymmetric boundary condition and obtain from it 
the corresponding 
%c4
mass deformed 
Nahm equation.

The Lorentzian 3-algebra is defined by a set of generators 
%c3 changes indices
$T^a=\{T^+, T^-, T^\ri \}$, 
where $T^\ri$ are the generators of a Lie algebra $\cG$ of the compact 
gauge group $G$ with the structure constant $f^{\ri \rj \rk}$ 
and Killing metric $\d^{\ri \rj}$.
%S4
The 3-bracket is specified by
\bea
\label{L3alg1}
\left[T^{-},T^a,T^b\right] &=& 0, \quad a=+,-,\ri, \\
\label{L3alg2} \left[T^+,T^\ri,T^\rj\right] &=& f^{\ri \rj}{}_\rk T^\rk, \\
\left[T^\ri,T^\rj,T^\rk\right] &=& f^{\ri \rj \rk}T^{-}.
\eea
%S4 and 
The invariant metric on this algebra is
\bea
\label{lmetric}
 \Tr (T^{-},T^{+}) &=& -1, \nn \\
 \Tr(T^{\ri},T^{\rj}) &=& \d^{\ri \rj}.
\eea
Expanding all the fields with respect to the generators
\be
X^I = X^I_a T^a = X^I_{-}T^{-} + X^I_+ T^+ + \Xh^I, 
\ee
where $\Xh^I = X^I_\ri T^\ri$ are the modes corresponds to the Lie algebra $\cG$,
one obtain the action for a Lorentzian BL theory \cite{gv},
\bea
\label{lorentz}
\cL_{Lorentz} = &&-\frac{1}{2}\Tr(\Dh_\m \Xh^I - B_\m X_+^I)^2 
+\del_\m X^I_+(\del_\m X_-^I-\Tr(B_\m, \Xh^I))
+\frac{1}{2}\varepsilon^{\m\n\l}\Tr(B_\l F_{\m\n})\nn\\
&&+\frac{i}{2}\Tr(\hat{\Psib}\Gamma^\m,(\Dh_\m\Psih -B_\m\Psi_+)) 
-\frac{i}{2}\hat{\Psi}_+\Gamma^\m(\del_\m\Psi_- -\Tr(B_\m,\Psih)) 
-\frac{i}{2}\Psi_-\G^\m\del_\m\Psi_+ 
\nn\\
&&+\frac{i}{2}\Tr(\hat{\Psib}\G^{IJ}X_+^I[\Xh^J,\Psih]) 
+\frac{i}{4}\Tr(\hat{\Psib}\G^{IJ}[\Xh^I,\Xh^J]\Psi_+) 
-\frac{i}{4}\Tr(\hat{\Psib}_+\G^{IJ}[\Xh^I,\Xh^J]\Psih)\nn\\
&&-\frac{1}{12}\Tr\left(X_+^I[\Xh^J,\Xh^K]+X_+^J[\Xh^K,\Xh^I] 
+ X_+^K[\Xh^I,\Xh^J]\right)^2,
\eea
where $I=3,\dots,10$. Here $A_\m$ is a gauge field for the compact gauge 
group $G$. The gauge field $B_\m$ is defined by
$B_\m = A_{\m \ri\rj}f^{\ri\rj}{}_\rk T^\rk$ and the theory is invariant under
an extra non-compact gauge symmetry associated with $B_\m$:
\bea \label{sgs}
\d B_\m &=& D_\m \z, \quad \d \Xh^I = \z X_+^I, 
\quad \d X_-^I = \Tr(\z,\Xh^I), \nn\\
\d \Psih &=& \z\Psi_+, \quad \d\Psi_- = \Tr(\z,\Psih).
\eea
The supersymmetry transformations read:
\bea
\label{d2susys}
\d_0 X_{-}^I &=&i\eb\Gamma^I\Psi_{-}, \quad \d_0 X_+^I 
=i\eb\Gamma^I\Psi_+, \quad \d_0\Xh^I = i\eb\Gamma^I\Psih, \\
\d_0 \Ah_\m &=& \frac{i}{2}\eb\G_\m\G^I(X_+^I\Psih - \Xh^I\Psi_+), \quad \d_0 B_\m 
= i\eb\G_\m\G^I[\Xh^I,\Psih], \\
\d_0\Psi_{-} &=& (\del_\m X_{-}^I 
-\Tr( B_\m X^I))\G^\m\G^I\e 
- \frac{1}{3}\Tr( \Xh^I\Xh^J\Xh^K ) \G^{IJK}\e, \nn \\ 
\d_0\Psi_+ &=& \del_\m X_+^I \G^\m\G^I\e, \\
\d_0\Psih &=& \Dh_\m \Xh^I \G^\m\G^I\e 
-\frac{1}{2}X_+^I[\Xh^J,\Xh^K]\G^{IJK}\e.
\eea

A special feature of the Lagrangian \eq{lorentz} 
is that the fields $X_-^I, \Psi_-$  
appear linearly. For convenience, let us collect the terms containing 
$X_-^I, \Psi_-$. It is
\bea
\cL_{gh} = \partial_\m X_+^I\partial^\m X_{-}^I 
-\Psib_{+}\G^\m \partial_\m\Psi_-.
\eea
We have called it a ghost term since $\cL_{gh}$ has an 
indefinite metric and is hence non-unitary. 
One can integrate out 
%c4 these fields 
$X^I_-, \Psi_-$ and obtain 
the equations of motion:
\bea
\label{eom1}\del^2X_+^I &=&0, \\
\label{eom2}\G^\m\del_\m\Psi_+ &=&0.
\eea
A solution to \eq{eom1} and \eq{eom2} which preserves gauge symmetry 
and supersymmetry is given by 
\bea
\label{vev1}
X_+^I &=& v_0\d_{10}^I, \\
\quad \Psi_+ &=& 0,
\eea
where $v_0\in \mathbb{R}$. Substituting this into the Lagrangian 
\eq{lorentz} and integrating out the field $B_\m$ gives us the Lagrangian
\bea
\label{n8sym}
\cL = &&-\frac{1}{2}(\Dh_\m \Xh^I)^2 
+ \frac{i}{2}\Tr(\hat{\Psib}, \G^\m D_\m \Psih) 
- \frac{1}{4v_0^2} \Tr(F_{\m\n},F^{\m\n}) 
-\frac{v_0^2}{4}\Tr([\Xh^I,\Xh^J],[\Xh^I,\Xh^J])\nn\\
&&+ \frac{iv_0}{2}\Tr(\hat{\Psib}\G^{(10)I}[\Xh^I,\Psih]) 
+\del_\l\left(\frac{\varepsilon^{\m\n\l}\Fh_{\m\n}\Xh^{10}}{2v_0}\right),
\eea
where we 
%c3 keep 
have kept the boundary term for later discussions. 
For a closed theory, the Lagrangian \eq{n8sym} is the maximally supersymmetric 
$\cN=8$ SYM theory in 2+1 dimensions. For an open theory, one will need an
appropriate boundary condition 
in order to decouple the field
$\Xh^{10}$  at the boundary. Moreover one gets additional boundary conditions
from requiring supersymmetry of the Lagrangian. 
For these boundary conditions
to be supersymmetric, $\e$ must be further restricted. 
This will be the subject of 
section \ref{sbc}.

We remark that 
apart from integrating out the fields $X_-^I, \Psi_-$, one can also keep them
and perform a BRST analysis by promoting a certain global shift symmetry  
to a local one \cite{schwarz,gv}. 
The analysis for the open case can be performed similarly.
In the following we will concentrate on the ``integrating out'' 
approach for our analysis.

\subsection{Multiple D2-branes in a background flux}

With the Lorentzian 3-algebra, the flux term and the mass term read
\bea
\label{Lf}
\cL_{flux} &=& 2\Gt_{IJKL} \Tr( X^I,[X^J,X^K,X^L]) \nn \\
&=& -8\Gt_{IJKL} X_+^I\Tr(\Xh^L, [\Xh^J,\Xh^K]), \\
\label{Lm}
\cL_{mass} &=& -\frac{1}{2}m^2 \Tr( X^I, X^I ) 
-\frac{i}{8}\Tr( \Psib,\G^{IJKL}\Psi) \Gt_{IJKL} \nn \\
&=& -\frac{1}{2}m^2 \Tr(\Xh^I, \Xh^I) +m^2 X_+^I X_{-}^I \nn\\
 &&-\frac{i}{8}\Tr(\hat{\Psib},\G^{IJKL}\Psih)\Gt_{IJKL} 
+\frac{i}{4}\Psib_{+}\G^{IJKL}\Psi_-\Gt_{IJKL}.
\eea
%c4 
As an aside, we note it is easy to check that the gauge symmetry \eq{sgs}
extends to the flux and mass Lagrangians \eq{Lf}, \eq{Lm}. We will take
$\mu =2m$ in the following analysis.

Due to the presence of new terms  linear in the fields $X_-^I$ and $\Psi_-$,
after integrating out these fields, we get the 
modified equations of motion
\bea
&&\partial^2 X_+^I -m^2 X_+^I = 0, \label{eom-X} \\
&&i\Gamma^\m\partial_\m\Psi_+ -\frac{i}{4}\G^{IJKL}\Psi_+\Gt_{IJKL} = 0.
\label{eom-Psi}
\eea
These are the Klein-Gordon and Dirac equations respectively. 
We will next show that one is able to 
pick solutions to $X_+^I$ and $\Psi_+$ which preserves gauge invariance 
and supersymmetry.

The supersymmetry transformations remains the same as for the bosons 
as in \eq{d2susys} but the flux modifies the supersymmetry 
transformations for the
fermions:
\bea
\label{susy-1f}
&& \d\Psi_{-} = (\del_\m X_{-}^I 
-\Tr( B_\m, X^I))\G^\m\G^I\e 
- \frac{1}{3}\Tr( \Xh^I\Xh^J\Xh^K ) \G^{IJK}\e 
+ \frac{1}{4}\Gts\G^M\e X_{-}^M,\qquad  \\ 
% && \qquad \qquad \qquad \qquad \qquad \qquad
% + \frac{1}{4}\G^{IJKL}\G^M\e X_{-}^M\Gt_{IJKL},\\
\label{susy+mode}
&& \d\Psi_+ = \del_\m X_+^I \G^\m\G^I\e 
+ \frac{1}{4}\Gts\G^M\e X_+^M, \\
\label{susyf}
&& \d\Psih = \Dh_\m \Xh^I \G^\m\G^I\e 
-\frac{1}{2}X_+^I[\Xh^J,\Xh^K]\G^{IJK}\e
+ \frac{1}{4}\Gts\G^M\e \Xh^M.
\eea
The simplest solution to \eq{eom-X} and \eq{eom-Psi} is
\bea
\label{psi+}
\Psi_+ &=& 0, \\
\label{coupling}
  X_+^I &=& 
  \begin{cases}
   v_0e^{m\s_1}\d_{10}^I,\ \ \  \text{or}\\
%S4 comma
   v_0e^{imt}\d_{10}^I ,
  \end{cases}
\eea
where 
$v_0$ is a real constant and the real part is assumed 
for the second solution in \eq{coupling}. 
As an illustration, we will consider below 
the first solution with the $\s_1$ dependence and for convenience we
will denote $v=v_0e^{m\s_1}$ below.
It is easy to see that the solution \eq{psi+} is supersymmetrically
invariant. In fact for the flux \eq{gen-G}, we have
%S4 changed \dt to delta  - now consistent with BL chapter. 
$\delta \Psi_+ = m v \G^{2(10)} ( 1- \G^2 R') \e$,
where $R'$ is defined by $R \G^{10} = \G^{(10)} R'$. Therefore the
configuration \eq{psi+} is supersymmetrically invariant for $\e$ satisfying 
\be \label{GRp}
(1-\G^{2} R')\e = 0.
\ee
Since the projectors 
\eq{c0p}, \eq{GRp} commute, 
8 supersymmetries are preserved.

%S4 added in eq refs to the solutions
Substituting the solution \eq{psi+}, \eq{coupling} and integrating 
out the $B_\mu$ fields, we finally obtain
%c3 I-> A
\bea
\label{md2}
\cL = &&-\frac{1}{2}(\Dh_\m \Xh^A)^2 
- \frac{1}{4v^2} \Tr(F_{\m\n},F^{\m\n}) 
-\frac{v^2}{4}\Tr([\Xh^A,\Xh^BB],[\Xh^A,\Xh^B]) \nn\\
&& + \frac{i}{2}\Tr(\hat{\Psib}, \G^\m D_\m \Psih) 
+ \frac{iv}{2}\Tr(\hat{\Psib},\G^{(10)A}[\Xh^A,\Psih]) \nn\\
&& - 8v \Gt_{(10)ABC}\Tr(\Xh^A,[\Xh^B,\Xh^C]) 
-\frac{i}{2}\Tr(\hat{\Psib},\G^{(10)ABC}\Psi)\Gt_{(10)ABC}
\nn\\
&& -\frac{1}{2}m^2\Tr(\Xh^A,\Xh^A) 
-\frac{i}{8}\Tr(\hat{\Psib},\G^{ABCD}\Psi)\Gt_{ABCD} \nn\\
&& +\del_\l\left(\frac{\varepsilon^{\m\n\l}}{2v}\Tr(\Fh_{\m\n},\Xh^{10})\right) 
-\frac{m}{2}\del_1\Tr(\Xh^{10},\Xh^{10}),
\eea
where the indices $A,B,C,D  =3, \cdots, 9$.
In the closed case, one can drop the last two terms in \eq{md2}.
Since the Lambert-Richmond action is supersymmetric and the 
solution \eq{coupling} is 1/2-BPS, by construction our action 
\eq{md2} is supersymmetric and preserves 8 supersymmetries:
%c3
\bea
\d \Xh^A &=& i \eb \G^A \Psih,\\
\d\Psih &=& \Dh_\m \Xh^A \G^\m\G^A\e 
-\frac{1}{2v} \e_{\m\n\l}\Fh^{\n\l} \G^\m \G^{10} \e
-\frac{v}{2}[\Xh^A,\Xh^B]\G^{AB(10)}\e
+ \frac{1}{4}\Gts\G^A\e \Xh^A,\quad \;\;\; \\
\d \Ah_\m &=& \frac{i v}{2} \eb \G_\m \G^{10} \Psib.
\eea

%S4 changed all IJKL indices to ABCD as they should
The Lagrangian \eq{md2} can be understood as the worldvolume theory of  
D2-branes with a space(time) dependent
coupling $g_{\rm YM}= v $ and coupled to NS-NS and R-R fluxes.
In 10 dimensions, the flux 
$\Gt_{ABCD}$ is identified with the R-R 4-form flux of 
the 3-form potential $C_3$ 
and $\Gt_{(10)ABC}$  is identified with the NS-NS 3-form flux of 
the 2-form potential $B_2$. 
%c2 
The term in \eq{md2} proportional to $\Gt_{(10)ABC}$ 
can be traced back 
as the low energy limit of the Myers' action \cite{myers}, 
together with its superpartner. 
The terms proportional to $m^2$ and $\Gt_{ABCD}$ are typical of 
%S4 couplings
couplings to the R-R fields. Supersymmetric Yang-Mills theories with a spacetime
dependent coupling were originally constructed in
%S Changed the title for cite for dzf as it came out strange in the biblio
 \cite{bak,dzf} 
and are known as Janus field theories. An extension to include a 
spacetime dependent
$\th$-angle for the 4-dimensional supersymmetric Yang-Mills 
was performed in \cite{CH} as an application to study spacetime 
singularities using holography. Similar field theory constructions also appear 
in the work \cite{witten}. 

In the open case, one needs to impose a boundary condition 
to decouple the $\Xh^{10}$ field at the
boundary. In particular we are interested in a supersymmetric boundary
condition in this paper.
This, together with the other boundary conditions that are
needed to maintain supersymmetry of the system, will be 
discussed next.
%c4 our next subject.

\subsection{Multiple D2-branes ending on a D4}
\label{sbc}

We now 
%c4 examine 
derive the supersymmetric boundary conditions for the flux
modified Lorentzian Bagger-Lambert theory. Since the 
field $X^I_-$ has been integrated out, the boundary condition \eq{r2} cannot be 
applied immediately and one needs to derive the boundary condition 
from the reduced action \eq{md2} directly.

Since rotational invariance is explicitly broken by the $\s_1$ 
dependence of the coupling, it makes a difference where the boundary is. 
For example,
the theory with a boundary at $\s_1=0$ is not equivalent to the theory 
with a boundary at $\s_2 =0$. In particular, to 
decouple the field $\Xh^{10}$ at the boundary, one needs to impose 
the boundary condition
\bea
\Tr(2\Fh_{02}\Xh^{10} +m v (\Xh^{10})^2) =0, \quad &&\mbox{boundary at $\s_1=0$}, 
\label{fbc1}
\\
\Tr(\Fh_{01} \Xh^{10}) =0, \quad && \mbox{boundary at $\s_2=0$}. \label{fbc2}
\eea
Let us first discuss the second case and assume that the condition \eq{fbc2}
is satisfied for the moment 
and come back to discuss whether it is supersymmetric later.

The supersymmetric variation of the Lagrangian is given by
%S changed l/rangles to Traces
%c3 change nottaion of susy
\be
\d \cL = \frac{i}{2}\partial_\m\Tr(\hat{\Psib},
\Gamma^\m\d\Psih) - \partial_\m\Tr(\d \Xh^A,
 \Dh^\m \Xh^A) +\mbox{bulk terms}.
\ee
Imposing the boundary condition $\s_2 =0$ gives 
\be \label{dL}
\d \int\mathrm{d}^3 \s \cL = \frac{i}{2} 
\int\mathrm{d}^2 \s\left( \Tr(\hat{\Psib},\Gamma^2\d\Psih)
-2 \Tr(\d X^A, \Dh_2 \Xh^A)\right),
\ee
so we obtain the boundary equation of motion
%c3 modified
\bea \label{bc-lorf}
\Dh_\m \Xh^A\hat{\Psib}\Gamma^2\Gamma^\m\Gamma^A\e 
 -\frac{v}{2}[\Xh^A,\Xh^B]\hat{\Psib}\Gamma^2\Gamma^{(10)AB}\e 
&& -\frac{1}{2v}\e_{\m\n\l}\Fh^{\n\l} \Psih \G^2 \G^{\m (10)} \e \\
&& +\frac{1}{4}\hat{\Psib}\Gamma^2 \Gts \Gamma^A\e \Xh^A 
-2\Dh_2\Xh^A\hat{\Psib}\Gamma^A\e = 0, \nn
\eea
where, $\m=0,1,2$.
%c3  as before and $A, B= 3,...,9$.
%c2 We will reduce to the $A,B$ index notation for the scalars at the
% end for simplicity.

Let us now consider a system of D2-branes 
ending on a D4-brane. In general a system of two intersecting D-branes is
supersymmetric if  
the relative transverse space has dimension in multiples of 4. 
Therefore, with  this in anticipation, let us look for a 
solution to the boundary condition \eq{bc-lorf} with 
%c4 X -> \Xh
\be \label{xzero}
\Xh^{3,4,5,6} =0.
\ee 
This corresponds to a D4-brane with worldvolume in the 01789-directions.
%S3
The R-symmetry is reduced from $SO(7)$ to $SO(3)$.
In addition to \eq{c0p}, 
we also impose the condition
\begin{equation}
\label{d4}
 \G^{01789(10)}\e = \e.
\end{equation}
%c4
We remark that this is not the same as the D4-brane projector in the 
$\kappa$-symmetric formulation of D-branes, see for example 
\cite{d1,d2}.
The effect of a background flux is already taken into 
%S4 account
account in terms of 
the M2-branes description and so the D4-brane supersymmetry is
represented simply by the condition
\eq{d4} in the M2-branes model.
%c4 We notice that the presence of $\G^{(10)}$ in \eq{d4} is as expected.
%c2 See for example \cite{bc} for a concise review of 
% the form of the supersymmetry projectors for D-branes.
From the above conditions, we obtain 
%c3 A,B -> i,j. use \Gt
\bea
\label{d2reln1}
%c6 %S4 swapped gamma order again
&&\G^{(10)ij}\e = - \varepsilon^{ijk}\G^2\G^k\e, \\
\label{d2reln2}&&\G^2\Gts \e = 4m\e,
\eea
where we have 
used the indices $i, j, k =7,8,9$.
%c3  and the indices $I,J,K,L = 3, \cdots, 10$.
%c3 We 
Let us also impose the condition
\be \label{psi-d4}
\G^{01789(10)} \Psi = -\Psi.
\ee
%c3 
It follows that $\delta \Ah_\a=0$ for $\a =0,1$ and hence
$\Fh_{01}$ is supersymmetric invariant. 
Therefore one can impose the supersymmetric 
boundary condition
\be
\Fh_{01} =0,
\ee
which also implies \eq{fbc2}.

The boundary condition \eq{bc-lorf} then simplifies to
\bea \label{nahm-d4}
%c6 %S4 sign on bracket term - could change it to a ... = 0 format equation.
\Dh_2\Xh^i = \frac{1}{2}v \varepsilon^{ijk}[\Xh^j,\Xh^k] +m\Xh^i.
\eea
The Nahm equation describes the profile of the 
D4-brane where the D2-branes end. The fuzzy funnel solution is obtained with 
the ansatz
\be \label{x-ansatz}
\Xh^i(\s_2)= f(\s_2) T^i,
\ee
where $T^i$ obey the $SU(2)$ algebra $[T^i,T^j] = \e^{ijk} T^k$ and $f$ obeys
\be
%c6 %S4 inserted sign on f^2 term
f'= vf^2 +mf.
\ee
This has solution
\be
%c6 %S4 modified the solution wrt the sign
f = \frac{m}{c e^{-m\s_2} - v},
\ee
where $c$ is a constant. The solution behaves as $f = v_0^{-1}/(s_0-\s_2)$ 
for small $m$, where $s_0$ is a constant. This is the expected profile in the 
absence of flux. In the presence of flux, the solution describes a fuzzy sphere
\be
\sum_{i=7}^{9}(\Xh^i)^2 = R^2,
\ee
whose  radius $R =C f$ depends on the Casimir $C$ of the 
representation as well as
$f$.  Since $v$ actually depends on $\s_1$, the fuzzy funnel has an $S^2$ 
cross section whose radius depends on both
$\s_1$ and $\s_2$. This is a new feature of the 
%c4 presence of flux.
flux we consider.

As before,  
%c3 one can check immediately 
it is straightforward to check that the boundary condition 
\eq{xzero}, \eq{psi-d4} and 
\eq{nahm-d4} are supersymmetric invariant. 
Finally we comment on the other possibility of having a boundary at $\s_1 =0$. 
Our analysis above can be performed in exactly the same way, with 
%c4 
only 
a  straightforward  change of the index 2 to 1 
in the equations \eq{dL}, \eq{bc-lorf}, \eq{d4}, 
\eq{d2reln1}, \eq{d2reln2}, \eq{psi-d4}.
%c4  we get the modified Nahm equation \eq{nahm-d4} with $D_2$ 
%c4 replaced by $D_1$. The solution will be more interesting also. 
However, it is easy to convince oneself that 
there is no way to impose a 
%S4 supersymmetry
supersymmetric boundary condition 
%c3 on $\Xh^{10}$ and $\Fh_{01}$ 
such that \eq{fbc1} holds. This is due to the
fact that $\Xh^{10}$ has a non-trivial supersymmetry variation. Therefore we 
conclude that with the solution $v =v_0 e^{m \s_1}$, 
the flux Lorentzian Bagger-Lambert theory is 1/2 BPS if there is a boundary 
at $\s_2 =0$. On the other hand, if the
boundary is at $\s_1 =0$, then all supersymmetries are broken.

\section*{Acknowledgements}
We are grateful to James Allen and Douglas Smith  for useful discussions. 
The work of CSC and GS is partially supported by STFC.

\end{document}